\documentclass[journal,onecolumn]{IEEEtran}

\ifCLASSINFOpdf

\else

\fi

\usepackage[T1]{fontenc}
\usepackage[linesnumbered,ruled]{algorithm2e}
\usepackage{stfloats,color}
\usepackage{amsfonts}
\usepackage{amssymb}
\usepackage[cmex10]{amsmath}
\usepackage{graphicx}
\usepackage{setspace}
\usepackage{cite}
\usepackage{array}
\usepackage{subcaption}
\usepackage{times}
\usepackage{epsfig}
\usepackage{latexsym}
\usepackage{epstopdf}
\usepackage{verbatim}
\usepackage{units}
\usepackage{amsthm}
\usepackage{placeins}
\usepackage{afterpage}
\usepackage{dsfont}
\usepackage{soul}
\usepackage{multicol}
\usepackage{multirow}
\usepackage{mathtools}
\usepackage[cmintegrals]{newtxmath}
\newcolumntype{P}[1]{>{\centering\arraybackslash}p{#1}}
\newcolumntype{M}[1]{>{\centering\arraybackslash}m{#1}}

\newcommand{\defeq}{\ensuremath{\triangleq}}

\newtheorem{theorem}{Theorem}

\newtheorem{remark}[theorem]{Remark}

\begin{document}

\title{Speeding Up Distributed Gradient Descent\\ by Utilizing Non-persistent Stragglers}

\author{\IEEEauthorblockN{Emre Ozfatura\IEEEauthorrefmark{2},
Deniz G\"und\"uz\IEEEauthorrefmark{2} and Sennur Ulukus\IEEEauthorrefmark{3}} \\
\IEEEauthorblockA{ \IEEEauthorrefmark{2}Information Processing and Communications Lab, Dept. of Electrical and Electronic Engineering,\\ Imperial College London, London, UK  \\
\IEEEauthorrefmark{3}Department of Electrical and Computer Engineering, Institute for Systems Research, \\ University of Maryland, College Park, MD \\
        {\tt \{m.ozfatura,d.gunduz\}@imperial.ac.uk, ulukus@umd.edu}
}\thanks{This work was supported by EC H2020-MSCA-ITN-2015 project SCAVENGE under grant number 675891, and by the European Research Council project BEACON under grant number 677854.}}


\maketitle

\begin{abstract}
When gradient descent (GD) is scaled to many parallel computing servers (CSs) for large scale machine learning problems, its per-iteration computation time is limited by the \textit{straggling} servers. Coded distributed GD (DGD) can tolerate straggling servers by assigning redundant computations to the CSs, but in most existing schemes, 
each non-straggling CS transmits one message per iteration to the aggregating server (AS) after completing all its computations. 
We allow multiple computations to be conveyed from each CS per iteration in order to exploit computations executed also by the straggling servers. We show that the average completion time per iteration can be reduced significantly at a reasonable increase in the communication load. We also propose a general coded DGD technique which can trade-off the average computation time with the communication load. 
\end{abstract}



\section{Introduction}

In many machine learning problems, for given $N$ training data points $\mathbf{X}=[\mathbf{x}_{1},\ldots,\mathbf{x}_{N}]^{T}$, $\mathbf{x}_i \in \mathbb{R}^{d}$, and the corresponding labels $\mathbf{y}=[y_{1},\ldots,y_{N}]^{T}$,  $y_{i}\in \mathbb{R}$, $i \in[N]\defeq \{1,2,\ldots,N\}$, the objective is to minimize the {\em parameterized empirical loss function}
\begin{equation}
L(\boldsymbol{\theta}) \triangleq \sum_{i=1}^{N}l\left((\mathbf{x}_{i}, y_{i}),\boldsymbol{\theta} \right) + \lambda R(\boldsymbol{\theta}),
\end{equation}
where $\boldsymbol{\theta}\in \mathbb{R}^{d}$ is the parameter vector, $l$ is an application specific loss function, and $R(\boldsymbol{\theta})$ is the regularization component. This optimization problem is commonly solved by gradient descent (GD), where at each iteration, the parameter vector $\boldsymbol{\theta}\in\mathbb{R}^{d}$ is updated along the GD direction:
\begin{equation}\label{update}
\boldsymbol{\theta}_{t+1} = \boldsymbol{\theta}_{t} - \eta_{t} \nabla_{\boldsymbol{\theta}} L(\boldsymbol{\theta}), ~
\end{equation}
where $\eta_{t}$ is the learning rate at iteration $t$,
and the gradient at the current parameter vector is given by $\nabla_{\boldsymbol{\theta}}=\sum_{i=1}^{N}\nabla_{\boldsymbol{\theta}}l\left((y_{i},x_{i}),\boldsymbol{\theta})\right)$.\\

When a large data set is considered, convergence of GD may take a long time, and distributed GD (DGD) techniques may be needed to speed up the convergence, where the computational task is divided into smaller sub-tasks and distributed across multiple computing servers (CSs) to be executed in parallel. In the beginning of the process, the aggregating server (AS) assigns $r$ sub-tasks to each CS, which may involve computing the gradient for $r$ different data points at each iteration. Whenever a CS completes  sub-tasks assigned to it, it sends  the results to the AS, where the results are aggregated to obtain $\boldsymbol{\theta}_{t+1}$, which is then transmitted to all the CSs to be used in the next iteration of the DGD algorithm. While distributed computation is essential to handle large data sets, the completion time of each iteration is constrained by the slowest server(s), called the \textit{straggling server(s)}, which can be detrimental for the convergence of the algorithm.\\
\indent Typically the computation and communication latency of CSs vary over time, and these values are not known in advance for a particular DGD session. The randomness of the persistent straggling servers can be considered to model a packet erasure communication channel, in which the transmitted data packets are randomly erased \cite{CC.1}. Motivated by this analogy, several papers have recently introduced coding theoretic ideas in order to mitigate the effect of straggling servers in DGD \cite{CC.1, UCCT.1, UCCT.3, UCUT.1}. The main idea behind these schemes is to introduce redundancy when allocating computation tasks to CSs in order to mitigate straggling servers.\\
\indent More recently, it has been shown that more efficient straggler mitigation techniques can be introduced  for specific computation tasks. particular attention has been paid to the least squares linear regression problem, which has the following loss function:
\begin{equation}
L(\boldsymbol{\theta}) = \frac{1}{2}\sum_{i=1}^{N}(y_{i}-\mathbf{x}_{i}^{T}\boldsymbol{\theta})^{2} ~.
\end{equation}
 For this particular model, the gradient is given by 
\begin{eqnarray}
\nabla_{\boldsymbol{\theta}} L(\boldsymbol{\theta}) = \mathbf{X}^{T} \mathbf{X} \boldsymbol{\theta}_{t}-\mathbf{X}^{T}\mathbf{y}.
\end{eqnarray}
Note that $\mathbf{X}^{T}\mathbf{y}$ remains the same throughout all the iterations, and the main computation task is to calculate $\mathbf{X}^{T}\mathbf{X}\boldsymbol{\theta}_{t}$. In this particular case the problem can be reduced to distributed matrix-matrix multiplication or matrix-vector multiplication, and the linearity of the gradient computation allows exploiting novel ideas from coding theory \cite{CC.1,CC.2,CC.3,CC.4,UNPS}.\\ 
 \indent Before the detailed explanation and analysis of these scheme we want to emphasize that in most of the straggling avoidance techniques designed for DGD, it is assumed that the straggling servers have no contribution to the computation task. However, in practice, \textit{non-persistent} straggling servers are capable of completing a certain portion of their assigned tasks. Therefore, our main objective in this paper is to redesign the straggling avoidance techniques in a way that computational capacity of the non-persistent stragglers can also be utilized. This will be achieved at the expense of an increase in the number of computations conveyed to the AS from the CSs, which we will define as the {\em communication load}.
 We first focus on the DGD scheme for the linear regression problem, then we consider another DGD strategy with uncoded computations, which can be applied to a general loss function.
\begin{table*}
\begin{center}
    \begin{tabular}{ | M{5cm} | M{5cm} |M{5cm} |}
    \hline
   UCUC & UCCC & CC\\ \hline
   \cite{UCUT.1}, \cite{UCUT.2}, \cite{UCUT.3} & \cite{UCCT.1}, \cite{UCCT.2}, \cite{UCCT.3} & \cite{CC.1} \cite{CC.2}, \cite{CC.3}, \cite{CC.4}\\ \hline
     \end{tabular}
 \end{center}
 \caption{Classification of the DGD algorithms in the literature according to the straggler avoidance approach used.}\label{table:coded_uncoded}
\end{table*}

\begin{table*}
\begin{center}
    \begin{tabular}{  | M{7.5cm} | M{7.5cm} |}
    \hline
 without pre-processing & with pre-processing \\ \hline
 \cite{CC.4}, \cite{UCCT.1} ,\cite{UCCT.2}, \cite{UCCT.3} & \cite{CC.1} ,\cite{CC.2} ,\cite{CC.3}, \cite{UNPS} \\ \hline
     \end{tabular}	
 \end{center}
 \caption{Classification of the DGD algorithms in the literature according to the application of  pre-processing on the data set.}\label{table:processing}
\end{table*}

\subsection{Straggler Avoidance Techniques}
In general, DGD schemes can be classified under three groups based on the employed straggling avoidance strategy; namely, 1) uncoded computation with uncoded communication (UCUC); 2) uncoded computation with coded communication (UCCC); and finally, 3) coded computation. The first group includes techniques in which the data points or mini-batches are distributed among the CSs, and each CS computes certain gradients, and returns results to the AS. In order to limit the completion time AS can update the parameter vector $\boldsymbol{\theta}_t$ after receiving only a limited number of gradients. The most common example of such schemes is  the stochastic gradient descent (SGD) approach with several different implementations, such as the K-sync SGD, K-batch-sync SGD, K-async SGD and K-batch-async SGD (see \cite{UCUT.1} for more details on these particular techniques). The schemes in the second group also distribute the data points in a similar fashion, but the computation results, i.e., values of the gradients, are sent to the AS in a coded form to achieve a certain tolerance against slow/straggling CSs \cite{UCCT.1,UCCT.2,UCCT.3}. While in uncoded computation the training data points are provided to the CSs as they are, in coded computation they are delivered in coded form \cite{CC.1,CC.2,CC.3,CC.4}. Classification of some of the DGD techniques in the literature into these three groups is given in Table \ref{table:coded_uncoded}. In all these schemes, the main idea is to assign redundant tasks to CSs in order to avoid straggling servers. We assume that $r$ tasks (these might correspond to $r$  data points or $r$ mini-batches depending on the application) assigned to each CS, which will be called the {\em computation load}.

In the \textit{gradient coding} approach \cite{UCCT.1}, a UCCC scheme, rows of $\mathbf{X}$, denoted by $\mathbf{x}_{1},\ldots,\mathbf{x}_{N}$ and which are also referred as data points,  are distributed to $N$ number of CSs\footnote{Throughout the paper, for simplicity, we assume that the number of data points are equal to the number of CSs, i.e, $N=K$, although the proposed schemes can be easily applied to any $N,K$ pair. Moreover, while we can refer data points, each data point can represent a mini batch of an arbitrary size depending on the application}.  Each row is assigned to multiple CSs to create redundancy. Each CS computes $\mathbf{x}\mathbf{x}^{T}\boldsymbol{\theta}_t$ for all the rows assigned to it, and sends a linear combination of these computations to the AS. In gradient coding the AS can recover the full gradient by receiving coded gradients from only $N-r+1$ CSs, at the expense of increased computation load at the CSs. Alternatively, in coded computation, linear combinations of the rows of $\mathbf{X}$ are distributed to CSs \cite{CC.4}. For each assigned coded input $\tilde{\mathbf{x}}$, the corresponding CS computes $\tilde{\mathbf{x}}\tilde {\mathbf{x}}^{T} \boldsymbol{\theta}_t$, and transmits the result to the AS. 

Note that $\mathbf{W}\defeq\mathbf{X}^{T}\mathbf{X}$ in $\nabla_{\boldsymbol{\theta}} L(\boldsymbol{\theta})$ remains the same throughout the iterations of the DGD process. Hence, if $\mathbf{W}$ is computed at the beginning of the process, the AS only requires the results of the inner products $\mathbf{w}_{1}^T\boldsymbol{\theta}_t,\ldots, \mathbf{w}_{N}^T \boldsymbol{\theta}_t$, where $\mathbf{w}_{i}$ is the $i$th row of $\mathbf{W}$. We call those schemes that work directly with data samples $\mathbf{X}$ as {\em distributed computation without preprocessing}, and schemes that work with $\mathbf{W}$ as  {\em distributed computation with preprocessing}. If $\mathbf{W}$ is available at the AS, the DGD for linear regression boils down to distributed matrix-vector multiplication, and the linear combinations of the rows $\mathbf{w}$ can be distributed to CSs as coded inputs \cite{CC.1,CC.2,CC.3,UNPS}. Classification of some of the known techniques in the literature according to pre-processing is given in Table \ref{table:processing}.

\subsection{Communication Load of DGD}

Coded computation and communication techniques are designed to ameliorate the effects of slow/straggling servers such that fast servers can compensate for the straggling ones. In most of the existing schemes, each non-straggling CS transmits a single message to the AS at each iteration of the DGD algorithm, conveying the results of all computation tasks assigned to it while the straggling servers do not transmit at all as they cannot complete their assigned tasks. This restriction leads to a trade-off between the per-server computation load, r, and the \textit{non-straggling threshold}, where the latter denotes the minimum number of CSs that must complete their tasks for the AS to recover all the gradients. This is achieved by assigning redundant computations to each of the CSs. In the extreme case, it may even be sufficient to get the results from only one CS, if all the computation tasks are assigned to each of the CSs, i.e., $r=N$.

However it is important to emphasize that a smaller non-straggling threshold does not necessarily imply a lower completion time; thus, the number of computations assigned to each CS and the non-straggling threshold should be chosen carefully. Indeed, beyond a threshold on the computation load $r$ (i.e., the number of computation tasks assigned to each CS), the average completion time starts increasing.

An important limitation of the  existing schemes in the literature is that the computations that have been carried out by the straggling servers are discarded, and not used by the AS at all; thus, the computation capacity of the network is underutilized. We show in this paper that the performance of the existing schemes can be improved by  allowing the communication of multiple messages from the CSs to the AS at each iteration of the employed DGD technique, so that CSs can send the results of partial computations before completing all the assigned computations at the expense of an increased \textit{communication load}, which characterizes the average number of total transmissions from the CSs to the AS per iteration. We remark that the overall impact of the increased communication load on the completion time depends on the distributed system architecture as well as the communication protocol used. The proposed multi-message techniques may be more attractive for special-purpose high performance computing (HPC) architectures employing message passing interface (MPI) rather than physically distributed machines communicating through standard networking protocols \cite{demistify:18}.


Multiple messages per server per iteration has recently been considered in \cite{CC.2} and \cite{UNPS}. In \cite{CC.2}, a hierarchical coded computation scheme is proposed, in which the computation tasks $\mathbf{w}_{1}\boldsymbol{\theta},\ldots,\mathbf{w}_{N}\boldsymbol{\theta}$ are divided into $L$ disjoint \textit{layers}. For each layer $l$ an $(n_{l},k_{l})$ MDS code is used for encoding the rows of $\mathbf{W}$, while the parameters $(n_{l},k_{l})$ are optimized according to the straggling statistics of the servers. Although this scheme provides an improvement compared to single-message schemes, it has two main limitations. First, the code design is highly dependent on the straggling behavior of the server, which is often not easy to predict, and can be time-varying. Second, if a sufficient number of coded computations for a particular layer are received to allow the decoding of the corresponding gradients, any further computations received for this particular layer will be useless. In that sense, a strategy with a single layer, i.e., $L=1$, will have a lower per iteration completion time when the decoding time is neglected. However, the decoding complexity at  AS is also affects the network performance, and this layered structure helps  reduce the decoding complexity. In \cite{UNPS}, the authors also consider the multi-message approach, but instead of using  MDS code with layered structure they use rateless codes, particularly LT codes, to reduce the decoding complexity. However, to achieve the introduced results, large number of coded messages should be passed to AS at each iteration, which induces the packetization problem that limits its applicability to real systems. 

\subsection{Objective and Contributions}

Although the aforementioned works \cite{CC.2,UNPS} allow multiple messages per server (per iteration), they assume the presence of a preprocessing step; that is, instead of the distribution of the rows of matrix $\mathbf{X}$ (or, their coded versions) as computation tasks, rows of matrix $\mathbf{W}$ are distributed.  However, obtaining $\mathbf{W}$ may not be practical for large data sets. Hence, we focus on the performance of coded computation and communication schemes that work directly on matrix $\mathbf{X}$, allowing multiple messages to be transmitted from each CS at each iteration. Moreover, in many scenarios with huge data sets, the data may not even be available centrally at the AS, and instead stored at the CSs to reduce the communication costs and the storage requirements at the AS. Therefore, we also consider uncoded computation techniques.

As we discussed previously, the schemes in the literature focus on minimizing the non-straggling threshold, which does not necessarily capture the average completion time statistics for one iteration of the GD algorithm. Indeed, in certain regimes of computation load $r$, the average completion time may be increasing as the non-straggling threshold decreases. Accordingly, in  this paper, we consider the average completion time as the main performance metric and develop DGD algorithms that can provide a trade-off between the communication load and the computation load.

To model the straggling behavior at the CSs, we use the model introduced in \cite{CC.1} to derive a closed form expression for the completion time statistics for both single and multi-message communication scenarios. We will also present numerical results based on Monte-Carlo simulations to compare the performances of different schemes in terms of the trade-off they obtain between the average completion time and the computation load. We also analyze the performance of an uncoded computation and communication scheme for the multi-message scenario, and show that in certain cases it outperforms its coded counterparts, while also significantly reducing the decoding complexity.

\section{Coded Computation}
We first explain the coded computation strategy  when there is no pre-processing step, i.e., $\mathbf{W}$ is not known in advance. For a given computational load constraint $r$, also called as the repetition factor, $r$ coded rows,  $\tilde{\mathbf{x}}_{i}^{(1)},\ldots,\tilde{\mathbf{x}}_{i}^{(r)}$ are assigned to $CS_{i}$ which executes the following computations $\tilde{\mathbf{x}}_{i}^{(1)}(\tilde{\mathbf{x}}_{i}^{(1)})^{T}\boldsymbol{\theta},\ldots,\tilde{\mathbf{x}}_{i}^{(r)}(\tilde{\mathbf{x}}_{i}^{(r)})^{T}\boldsymbol{\theta}$. Once all these computations are executed, $CS_{i}$ returns their sum to AS. The results obtained from a sufficient number of CSs are used at the AS to compute the next iteration of the parameter vector, $\boldsymbol{\theta}_{t+1}$. Now we will briefly summarize the Lagrange coded computation method introduced in \cite{langrange,CC.4}, which utilizes polynomial interpolation for the code design. 
\subsection{Lagrange Polynomial}
Consider the following polynomial
\begin{equation}
f(z)\defeq\sum_{i\in[N]}\mathbf{a}_{i}\prod_{j\in[N]\setminus\left\{i\right\}} \frac{z-\alpha_{j}}{\alpha_{i}-\alpha_{j}},
\end{equation}
where $\alpha_{1},\ldots,\alpha_{N}$ are $N$ distinct real numbers, and $\mathbf{a}_{1},\ldots,\mathbf{a}_{N}$ are vectors of size $1\times k$. The main feature of the $f(z)$ polynomial is that; $f(\alpha_{i})=\mathbf{a}_{i}$, for $i\in[N]$. Let us consider another polynomial 
\begin{equation}
h(z)=f(z)f(z)^{T}\boldsymbol{\theta},
\end{equation}
such that\footnote{We dropped the time index on  $\boldsymbol{\theta}$ for brevity.} $h(\alpha_{i})=\mathbf{a}_{i}\mathbf{a}_{i}^{T}\boldsymbol{\theta}$. Hence, if the coefficients of  polynomial $h(z)$ are known, then the term $\sum_{i=1}^{N}\mathbf{a}_{i}\mathbf{a}_{i}^{T}\boldsymbol{\theta}_t$ can be obtained easily. We remark that the degree of the polynomials $f(z)$ and $h(z)$ are $N-1$ and $2N-2$, respectively. Accordingly, 
if the value of $h(z)$ at $2N-1$ distinct points  are known at the AS, then all its coefficients can be obtained via polynomial interpolation. This is the key notion behind  Lagrange coded computation, which is explained in the next subsection.

\subsection{Lagrange Coded Computation (LCC)}

Let us first assume that $N$ is multiple of $r$ For given $r$ and $N$, the rows of $\mathbf{X}$, $\mathbf{x}_{1},\ldots,\mathbf{x}_{N}$, are divided into $r$ disjoint groups, each of size $N/r$, and the rows within each group are ordered according to their indices. Let $\mathbf{x}_{k,j}$ denote the $j$th row in the $k$th group, and $\mathbf{X}_{k}$ denote all the rows in the $k$th group; that is, $\mathbf{X}_{k}$ is the $N/r \times d$ submatrix of $\mathbf{X}$. Then, for distinct real numbers $\alpha_1, \ldots, \alpha_{N/r}$, we form the following $r$ structurally identical polynomials of degree $N/r-1$ take the rows of $\mathbf{X}_{k}$ as their coefficients:
\begin{equation}
f_{k}(z)=\sum_{i=1}^{N/r}\mathbf{x}_{k,i}\prod_{j=1,j \neq i}^{N/r} \frac{z-\alpha_{j}}{\alpha_{i}-\alpha_{j}},\text{ } k\in[r].
\end{equation}
Then we define
\begin{equation}
 H(z)\defeq\sum_{k=1}^{r}f_{k}(z)f_{k}(z)^{T}\boldsymbol{\theta}_t.
\end{equation}
Coded vectors $\tilde{\mathbf{x}}_{i}^{(k)}$, $k\in[r]$, for $CS_{i}$, $i\in[N]$ are obtained by evaluating
$f_{k}(z)$ polynomials  at distinct values, $\beta_{i} \in \mathbb{R}$, i.e.,
$\tilde{\mathbf{x}}_{i}^{(k)} = f(\beta_{i})$. 
At each iteration of the DGD algorithm $CS_{i}$ returns the value of 
 \begin{equation}
 H(\beta_{i})=\sum_{k=1}^{r}\tilde{\mathbf{x}}_{i}^{(k)}(\tilde{\mathbf{x}}_{i}^{(k)})^{T}\boldsymbol{\theta}_t.
 \end{equation}
The degree of polynomial $H(z)$ is $2N/r-2$; and thus, the non-straggling threshold for LCC is given by $K_{LCC}(r)=2N/r-1$; that is, having received the value of $H(z)$ at $K_{LCC}(r)$ distinct points, the AS can extrapolate $ H(z)$ and compute
 \begin{equation}
 \sum_{j=1}^{N/r}H(\alpha_{j})=\mathbf{X}^{T}\mathbf{X}\boldsymbol{\theta}_t,
 \end{equation}
\indent When $N$ is not divisible by $r$, zero-valued data points can be added to $\mathbf{X}$ to make it divisible by $r$. Hence, in general the non-straggling threshold is given by $K_{LCC}(r)=2\lceil N/r \rceil-1$. 
\subsection{LCC with Multi-Message Communication}
LCC for distributed gradient descent has been originally proposed in \cite{langrange,CC.4} considering the transmission of a single-message to the AS per CS per iteration. Here, we introduce a multi-message version of LCC by using a single polynomial $f(z)$ of degree $N-1$, instead of using $r$ different polynomials, each of degree $N/r-1$. We define
\begin{equation}
f(z)\defeq\sum_{i=1}^{N}\mathbf{x}_{i}\prod_{j=1,j\neq i}^{N} \frac{z-\alpha_{j}}{\alpha_{i}-\alpha_{j}},
\end{equation}
where $\alpha_{1},\ldots,\alpha_{N}$ are  $N$ distinct real numbers, and we construct
\begin{equation}
h(z)\defeq f(z)f(z)^{T}\boldsymbol{\theta}_t,
\end{equation}
such that $h(\alpha_{i})=\mathbf{x}_{i}\mathbf{x}_{i}^{T}\boldsymbol{\theta}_t$. Consequently, if the polynomial $h(z)$ is known at the AS, then the full gradient $\sum_{i=1}^{N}h(\alpha_{i})=\sum_{i=1}^{N}\mathbf{x}_{i}\mathbf{x}_{i}^{T}\boldsymbol{\theta}_{t}$ can be obtained. To this end, $r$ coded vectors $\tilde{\mathbf{x}}_{i}^{(1)},\ldots,\tilde{\mathbf{x}}_{i}^{(r)}$, which are assigned to $CS_{i}$, $i\in[N]$ are constructed by evaluating $f(z)$ at $r$ different points, $\beta_{i}^{(1)},\ldots,\beta_{i}^{(r)}$, i.e.,
\begin{equation}
\tilde{\mathbf{x}}_{i}^{(j)}=f(\beta_{i}^{(j)}), \text{ }i\in[N], j\in[r].
\end{equation}

$CS_{i}$ computes $\tilde{\mathbf{x}}_{i}^{(1)}(\tilde{\mathbf{x}}_{i}^{(1)})^{T}\boldsymbol{\theta}_t,\ldots,\tilde{\mathbf{x}}_{i}^{(r)}(\tilde{\mathbf{x}}_{i}^{(r)})^{T}\boldsymbol{\theta}_t$, and transmits the resultant vector to the AS after each computation. Coded computation corresponding to coded data point $\tilde{\mathbf{x}}_{i}^{(j)}$ at $CS_{i}$ provides the value of polynomial $h(z)$ at point $\beta_{i}^{(j)}$. The degree of the polynomials $f(z)$ and $h(z)$ are $N-1$ and $2(N-1)$, respectively, which implies that  $h(z)$ can be interpolated from its values at any $2N-1$ distinct points. Hence, any $2N-1$ computations received from any subset of the CSs are sufficient to obtain the full gradient.\\
\indent We note that, in the original LCC scheme coded data points are constructed evaluating $r$ different polynomials at the same data point, whereas in the multi-message LCC scheme, coded data points are constructed evaluating a single polynomial at $r$ distinct points. In multi-message scenario, per iteration completion time can be reduced since the partial computations of the non-persistent stragglers are also utilized; however, at the expense of an increase in the communication load. Nevertheless, it is possible to set the number of polynomials to a different value to seek balance between the communication load and the per iteration completion time. This will be illustrated in Section \ref{s:numerical_results}. 

\section{Uncoded Computation and Communication (UCUC)}\label{s:UCUC}

In UCUC, the data points are divided into $N$ groups, where $N$ is the number of CSs, and each group is assigned to a different CS. While the per iteration completion time is determined by the slowest CS in this case, it can be reduced by assigning multiple data points to each CS, and allowing it to communicate the result of its computation for each data point right after its execution. We note here that, with UCUC the AS can apply SGD, and evaluate the next iteration of the parameter vector without waiting for all the computations. While we will mainly consider GD with a full gradient computation in our analysis for a fair comparison with the presented CGD approaches, we will show in Section \ref{s:numerical_results} that significant gains can be obtained in both computation time and communication load by ignoring only 5\% of the computations.

Let $\mathbf{A}$ be the assignment matrix for the data points to CSs, where $\mathbf{A}(j,k)=i$  means that the $i$th data point is computed by the $k$th CS in the $j$th order.\\ 
\indent An easy and efficient way of  constructing $\mathbf{A}$ is to use a circular shift matrix, where
\begin{equation}
\mathbf{A}(j,:)=\text{circshift }([1:N],-(j-1)).
\end{equation}
 For instance, for $N=K=10$ and $r=4$, we have:
\[
\mathbf{A}=
  \begin{bmatrix}
    1 & 2 & 3 & 4 & 5 & 6 & 7 & 8 & 9 & 10\\
    2 & 3 & 4 & 5 & 6 & 7 & 8 & 9 & 10 & 1 \\
    3 & 4 & 5 & 6 & 7 & 8 & 9 & 10 & 1 & 2  \\
    4 & 5 & 6 & 7 & 8 & 9 & 10 & 1 & 2 & 3\\
  \end{bmatrix}.
\]
We highlight that, in the multi-message scenario uncoded communication always outperforms the gradient coding scheme of \cite{UCCT.1}. In the latter, a necessary condition to obtain the full gradient is that each partial gradient, i.e., the gradient corresponding to one data point, is computed by at least one server. It is easy to see that, under this condition, full gradient can also be obtained by UCUC. Hence, the main advantage of the  gradient coding scheme is to minimize the communication overhead. Hence, we do not consider a multi-message gradient coding scheme. We note here that the utilization of the non-persistent stragglers in the single-message UCUC scenario is studied in \cite{UCUT.3}. In the scheme proposed in \cite{UCUT.3}, instead of sending each gradient separately, each CS transmits the sum of the gradients computed up until a specified time constraint, and, these sums are combined at the AS using different weights.
\begin{figure*}
    \centering
         \begin{subfigure}[b]{0.47\textwidth}
        \includegraphics[scale=0.6]{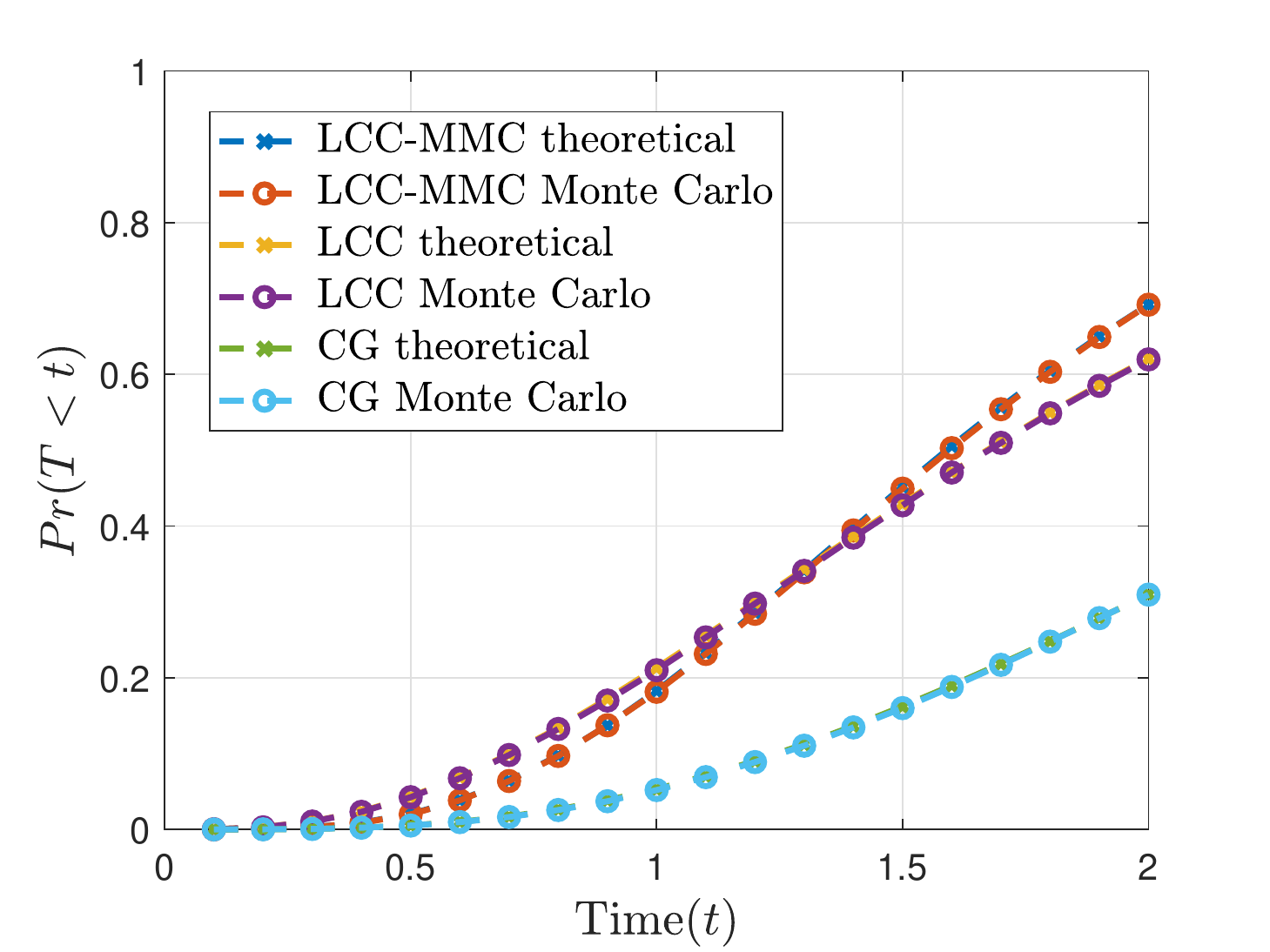}
        \caption{$N=6$, $r=3$}
				\label{N6}
    \end{subfigure}
    \begin{subfigure}[b]{0.47\textwidth}
        \includegraphics[scale=0.6]{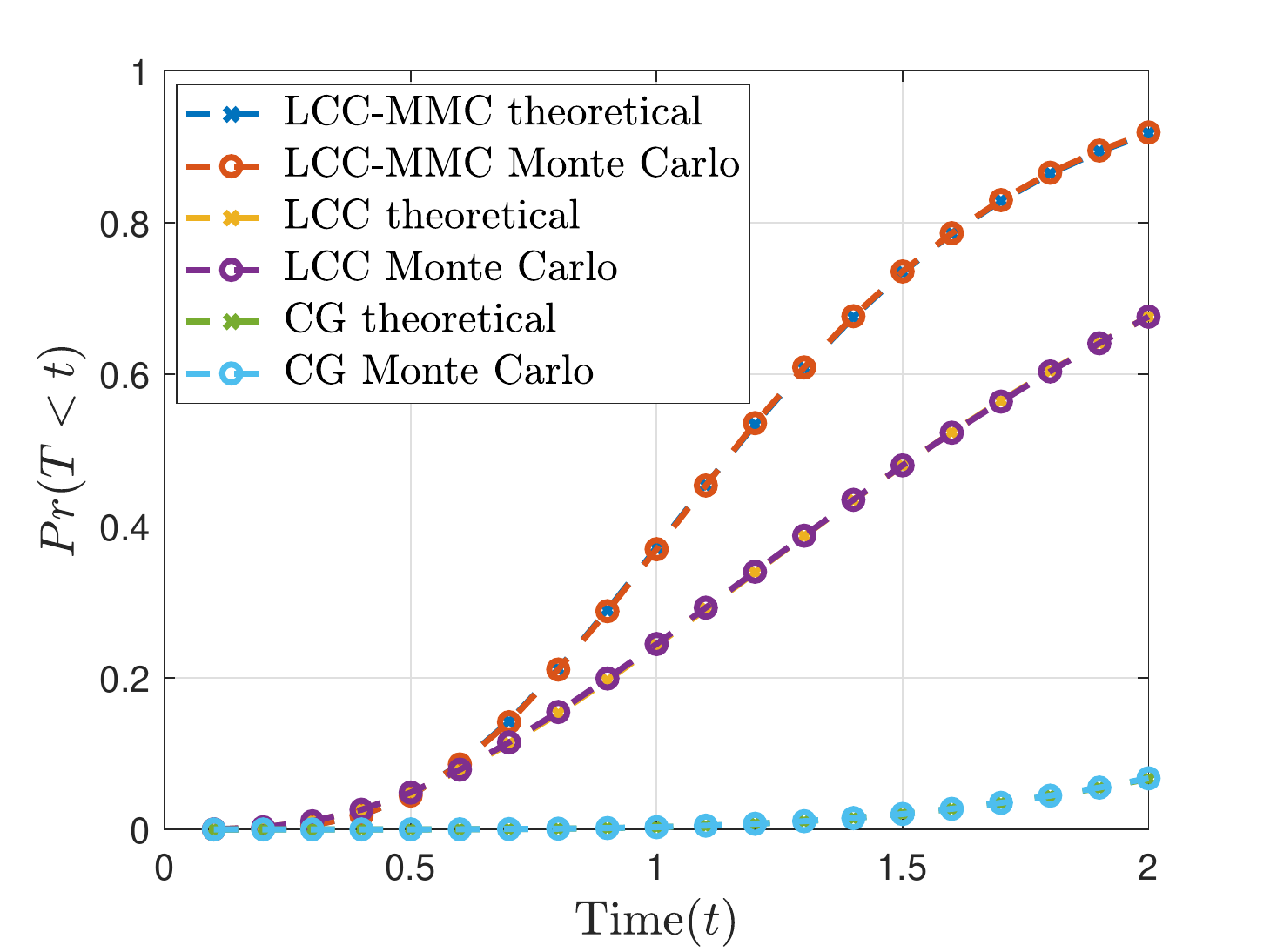}
        \caption{$N=10$, $r=5$}
				\label{N10}
        \end{subfigure}
				\caption{Per iteration completion time statistics. }
		\label{res1}
\end{figure*}

\section{Per Iteration Completion Time Statistics}

In this section, we analyze the statistics of  per iteration completion time $T$ for the DGD schemes introduced above. For the analysis we consider a setup with $N$ CSs and similarly we assume that the data set is divided into $N$ data points. For the straggling behavior, we adopt the model in \cite{CC.1} and \cite{CC.2}, and assume that the probability of completing $s$ computations at any server, such as multiplying $\boldsymbol{\theta}$ with $s$ different coded rows $\tilde{\mathbf{w}}$, by time $t$ is given by
\begin{equation}\label{dist}
F_{s}(t)\defeq
    \begin{cases}
     1-e^{-\mu(\frac{t}{s}-\alpha)}, &  \text{if } t\geq s\alpha, \\
      0,   &  else. 
    \end{cases}
\end{equation}
The statistical model considered above is a shifted exponential distribution, such that the duration of a computation cannot be less than $\alpha$. We also note that, although the overall computation time at a particular CS has an exponential distribution, the duration of  each computation is assumed to be identical. Further, let $P_{s}(t)$ denote the probability of completing exactly $s$  computations by time $t$. We have
\begin{equation}
F_{s}(t)=\sum_{s^{\prime}=s}^{r}P_{s^{\prime}}(t),\label{corr}
\end{equation}
where $P_{r}(t)=F_{r}(t)$, since there are a total of $r$ computations assigned to each user. One can observe from (\ref{corr}) that  $P_{s}(t)=F_{s}(t)-F_{s+1}(t)$; and, hence $P_{s}(t)$ can be written as follows:
\begin{equation}
P_{s}(t)= 
    \begin{cases}
     0, &  \text{if } t<s\alpha, \\
     1-  e^{-\mu(\frac{t}{s}-\alpha)} ,  & s\alpha \leq t <(s+1) \alpha.\\
           e^{-\mu(\frac{t}{s+1}-\alpha)}-e^{-\mu(\frac{t}{s}-\alpha)},   &(s+1)\alpha<t, 
     \end{cases}
\end{equation} 

We divide the CSs into $r+1$ groups according to the number of computations completed by time $t$. Let $N_{s}(t)$ be the number of CSs that have completed  exactly  $s$  computations by time $t$, $s = 0, \ldots, r$, and define $\mathbf{N}(t) \triangleq (N_{0}(t),\ldots,N_{r}(t))$, where $\sum_{s=0}^{r}N_{s}(t)=N$. The  probability of a particular realization is given by
\begin{equation}
\mathrm{Pr}(\mathbf{N}(t))=\prod_{s=0}^{r} P_{s}(t)^{N_{s}}{N-\sum_{j<s}N_{j}\choose N_{s}}.
\end{equation}
At this point, we introduce $M(t)$, which denotes the total number of computations completed by all the CSs by time $t$, i.e., $M(t)\defeq\sum_{s=1}^{r}s \times N_{s}(t)$, and let $M_{th}$  denote the threshold for obtaining the full gradient\footnote{Recall that this threshold is either $N$ or $2N-1$ depending on the existence of a preprocessing step.}. Hence, the probability of recovering the full gradient at AS
by time $t$, $\mathrm{Pr}(T \leq t)$, is given by  $\mathrm{Pr}(M(t) \geq M_{th})$. Consequently, we have
\begin{equation}\label{stat1}
\mathrm{Pr}(T \leq t)=\sum_{\mathbf{N}(t):M(t)\geq M_{th}} \mathrm{Pr}(\mathbf{N}(t)),
\end{equation}
and
\begin{align}
E[T] & = \int_0^\infty \mathrm{Pr}(T > t) dt\\ 
& =\int_0^\infty \left[1 - \sum_{\mathbf{N}(t):M(t)\geq M_{th}} \mathrm{Pr}(\mathbf{N}(t)) \right] dt.
\end{align}

Per iteration completion time statistics of non-straggler threshold based schemes can be derived similarly. For a given non-straggler threshold $K_{th}$, and per server computation load $r$, we can have
\begin{equation}\label{stat2}
\mathrm{Pr}(T \leq t)=\sum_{k=K_{th}}^{N} {N \choose k}(1-e^{-\mu(\frac{t}{r}-\alpha)})^{k}(e^{-\mu(\frac{t}{r}-\alpha)})^{N-k},
\end{equation}
when $t\geq r \alpha$, and $0$ otherwise.

\section{Numerical Results}\label{s:numerical_results}

We first verify the correctness of the expressions provided for the per iteration completion time statistics in (\ref{stat1}) and  (\ref{stat2}) through Monte Carlo simulations  generating 100000 independent realizations. Then, we will show that the multiple-message communication approach can reduce the average per-iteration completion time $E[T]$ significantly. In particular, we analyze the per iteration completion time of different DGD schemes, coded gradient (CG), Lagrange coded computation (LCC),  and LCC with multi-message communication (LCC-MMC). For the simulations we consider two different settings, with $K=N=6$, $r=3$ and $K=N=10$, $r=5$, respectively, and  use the cumulative density function (CDF) in (\ref{dist}) with parameters $\mu=10$ and $\alpha=0.01$ for the completion time statistics.\\
\indent  In Fig.\ref{res1} we plot the CDF of the per iteration completion time $T$ for CG, LCC,  and LCC-MMC schemes according to the closed form expressions derived in Section 4 and Monte Carlo simulations. We observe from Fig. \ref{res1} that the provided closed-form expressions match perfectly with the results from the Monte Carlo simulations.  We also observe that, although the LCC-MMC and LCC schemes perform closely in the first scenario, LCC-MMC outperforms the LCC scheme in the second scenario. This is because, when the per user computation load $r$ is increased, it will take more time for even the fast CSs to complete all the assigned computations, which results in a higher number of non-persistent stragglers. Hence, the performance gap between LCC-MMC and LCC increases with $r$. Similarly, we also observe that CG performs better for small $r$ when the $N/r$ ratio is preserved.

 \begin{figure}
    \centering        \includegraphics[scale=0.6]{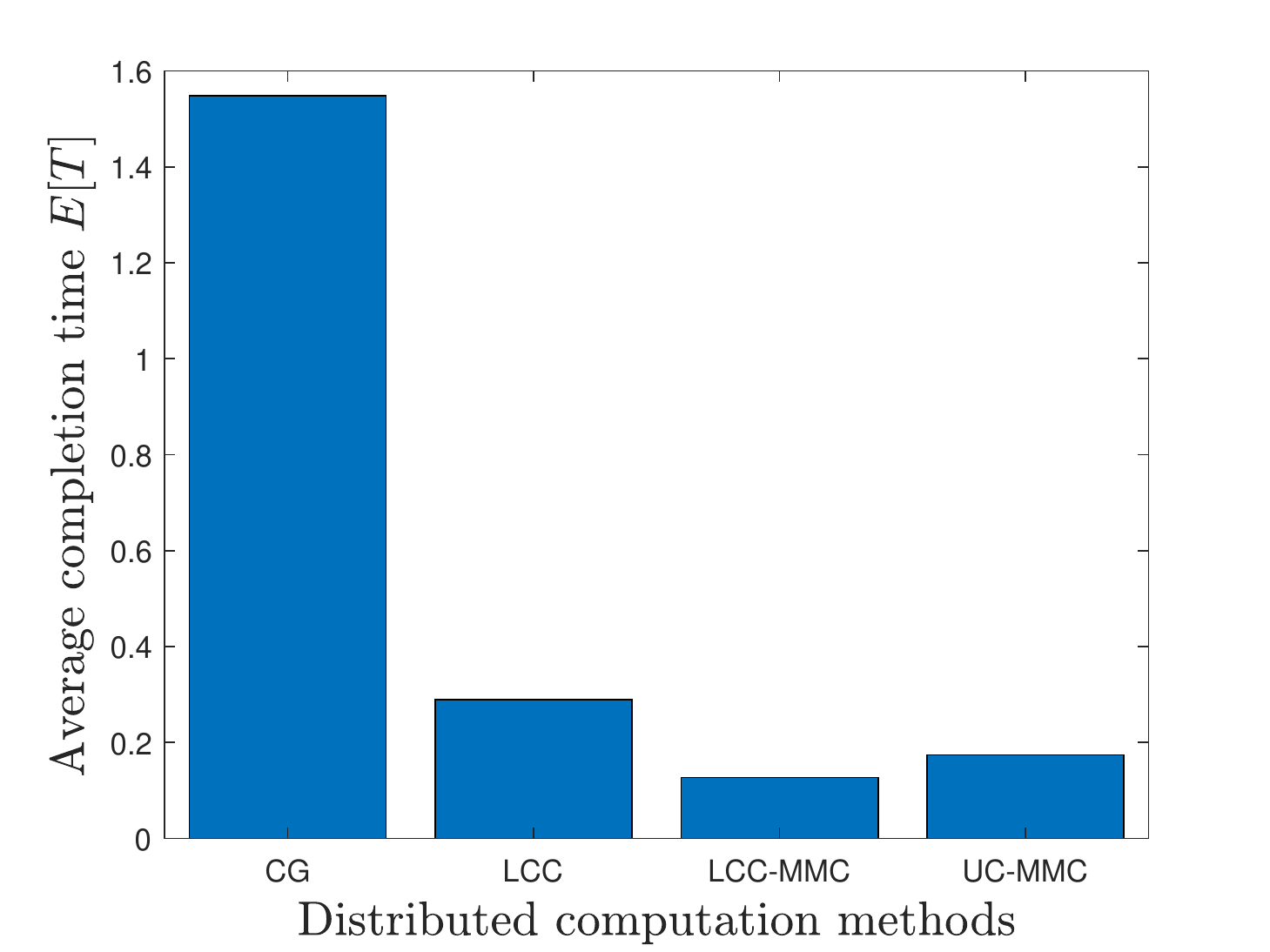}				\caption{Average completion time per iteration for $K=N=40$ and $r=10$.}\label{avgtime}
\end{figure}
\begin{figure*}
    \centering
         \begin{subfigure}[b]{0.47\textwidth}
        \includegraphics[scale=0.6]{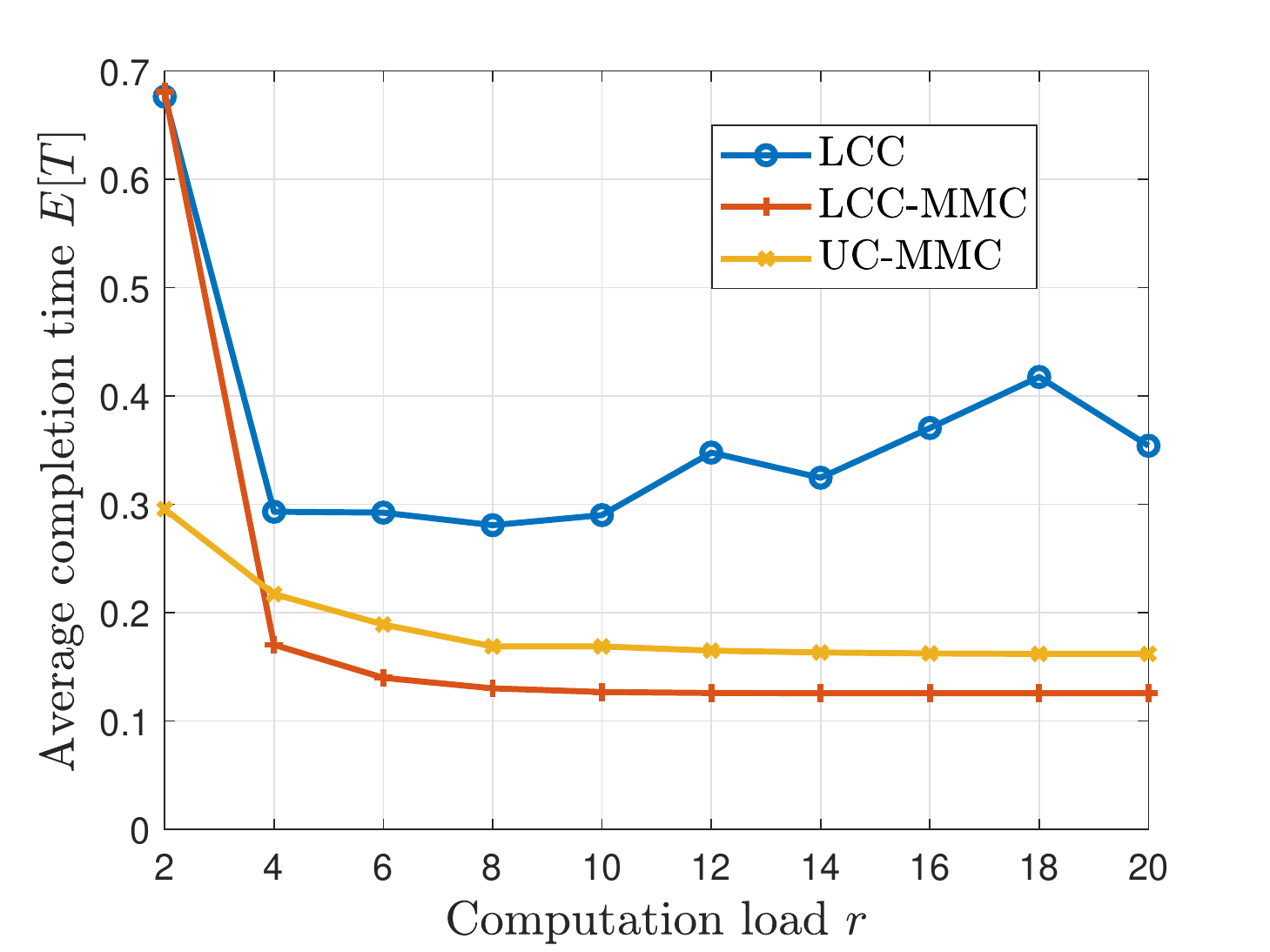}
        \caption{Average completion time vs. computation load.}
				\label{comp}
    \end{subfigure}
    \begin{subfigure}[b]{0.47\textwidth}
        \includegraphics[scale=0.6]{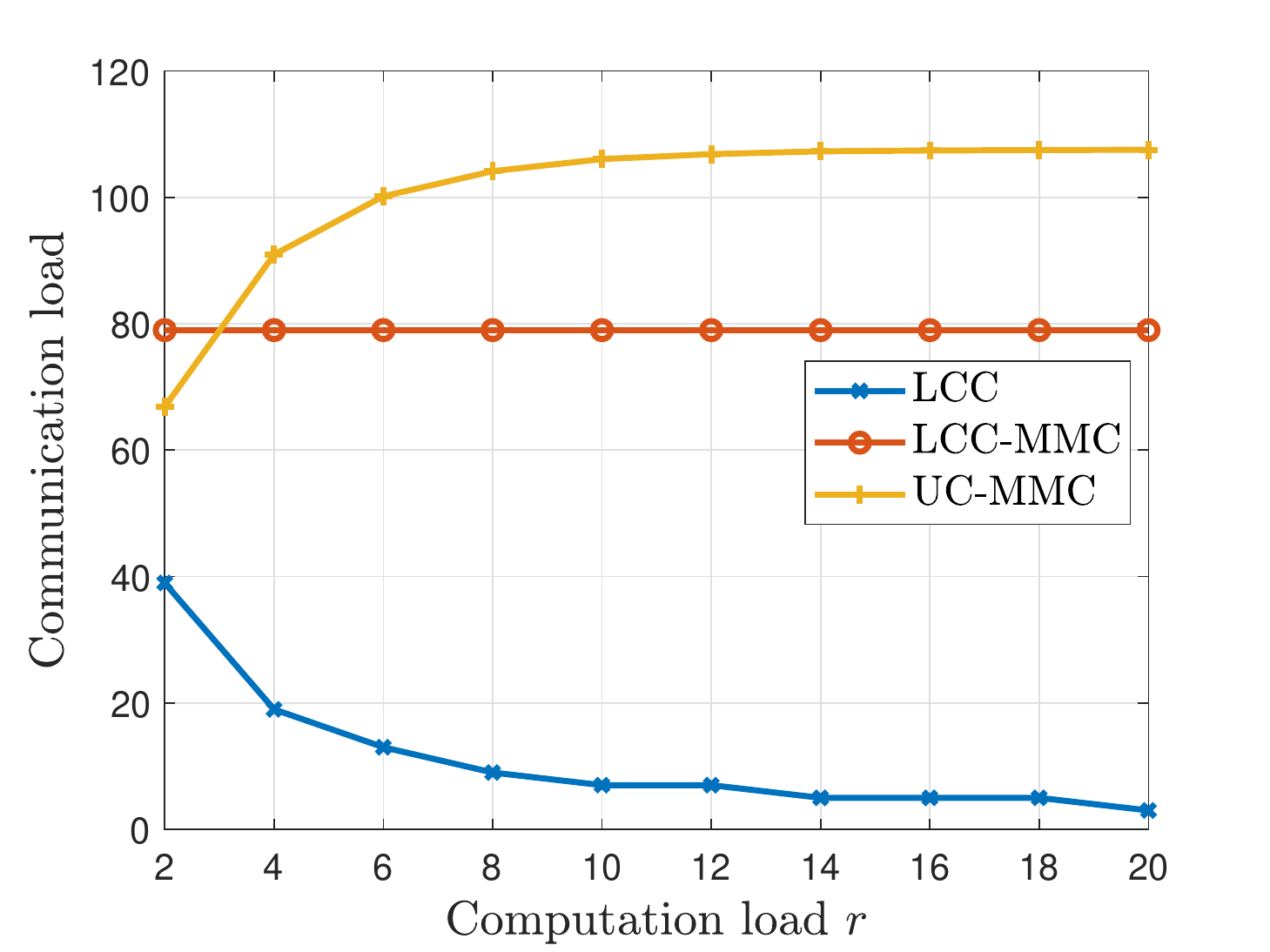}
        \caption{Communication load vs. computation load.}
				\label{comm}
        \end{subfigure}
				\caption{Per iteration completion time and communication load statistics.}
		\label{avg_numbtrans}
\end{figure*}

Next, we consider the setup from \cite{CC.4}, where $N=40$ CSs are assigned $K=40$ tasks to be computed at each iteration, where $r=10$ different computations are assigned to each server. Again, we use the distribution in (\ref{dist}) with parameters $\mu=10$ and $\alpha=0.01$. We compare the average per iteration completion time, $E[T]$, of the CG, LCC and LCC-MMC schemes, as well as the uncoded scheme with multi-message communication, UC-MMC, and the results are illustrated in Fig. \ref{avgtime}. We observe that  LCC-MMC approach can provide approximately $50\%$ reduction in the average completion time compared to  LCC, and more than $90\%$ reduction compared to GC. A more interesting result is that the UC-MMC scheme outperforms both LCC and GC. This result is especially important since UC-MMC has no decoding complexity at the AS. Hence, when the decoding time of AS is also included in the average per iteration completion time this improvement will be even more significant.

Finally, we analyze the performance of the various DGD schemes with respect to computation load $r$. We consider the previous setup with $N=K=40$, and consider $r=2,4,\ldots,20$. For the performance analysis, we consider both  the average per iteration completion time $E[T]$ and the communication load, measured by the average total number of transmissions from the CSs to the AS, and the results obtained from $100000$ Monte Carlo realizations are illustrated in Fig. \ref{avg_numbtrans}. From Fig. \ref{avg_numbtrans}(a), we observe that the UC-MMC scheme consistently outperforms LCC for all computation load values. More interestingly, UC-MMC performs very close to LCC-MMC, and for a  small $r$, such as $r=2$, it can even outperform UC-MMC. Hence, in terms of the computation load UC-MMC can be considered as a better option compared to  LCC especially when $r$ is low.\\
\indent On the other hand, from Fig. \ref{avg_numbtrans}(b) we observe that, in terms of the communication load the best scheme is LCC, while the UC-MMC introduces the highest communication load. We also  observe that communication load of the LCC-MMC scheme remains constant with $r$, whereas that of the LCC (UC-MMC) scheme monotonically decreases (increases) with $r$. Accordingly,  the communication load of the LCC and UC-MMC schemes are closest at $r=2$. From both  
Fig. \ref{avg_numbtrans}(a) and Fig. \ref{avg_numbtrans}(b) we note that, when $r$ is low, e.g., when the CSs have small storage capacity, UC-MMC may outperform the LCC scheme in terms of the average per iteration completion time including the decoding time as well.

\begin{figure*}[t]
    \centering
         \begin{subfigure}[b]{0.47\textwidth}
        \includegraphics[scale=0.6]{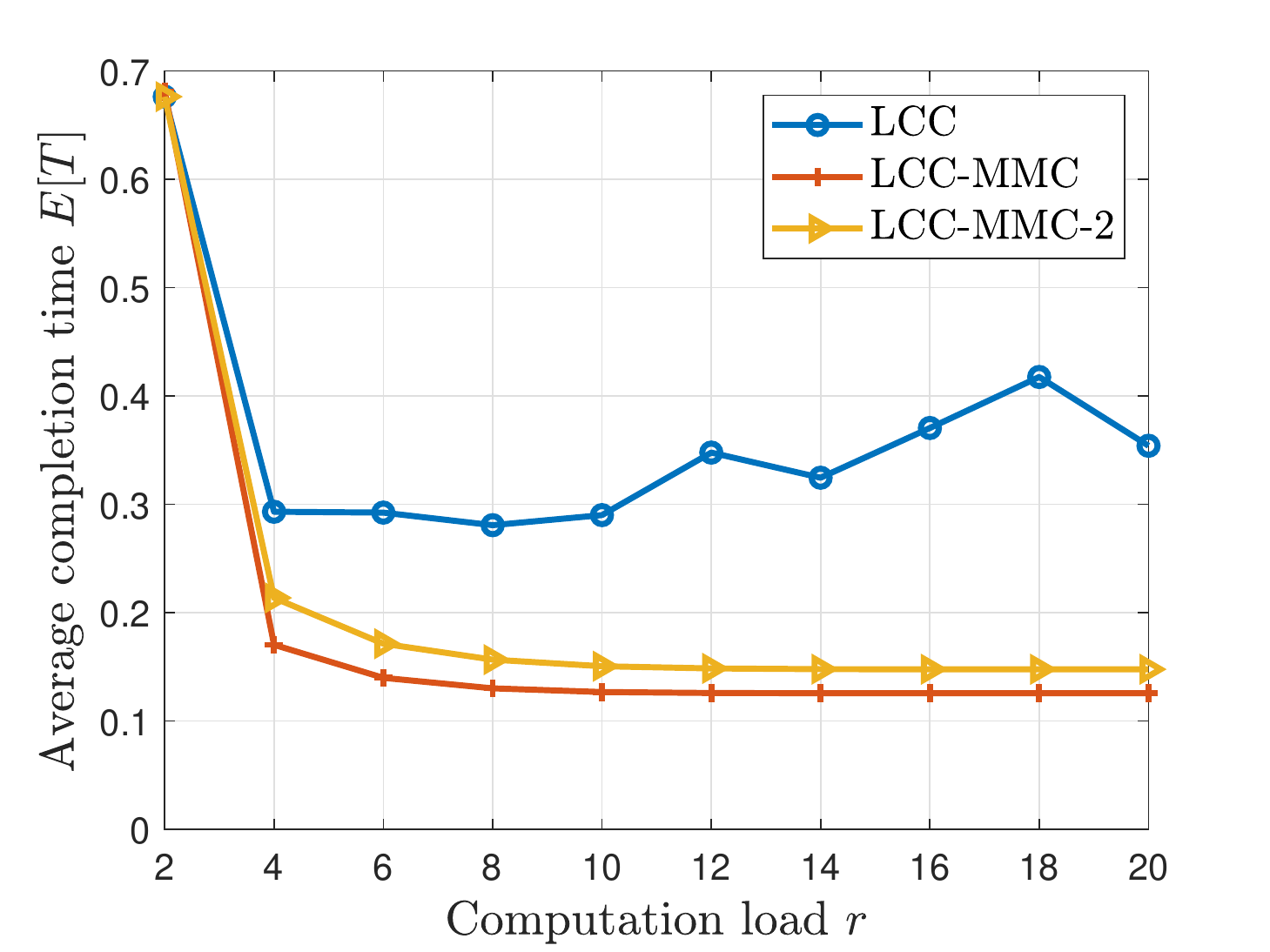}
        \caption{Average completion time vs computation load}
				\label{comp2}
    \end{subfigure}
    \begin{subfigure}[b]{0.47\textwidth}
        \includegraphics[scale=0.6]{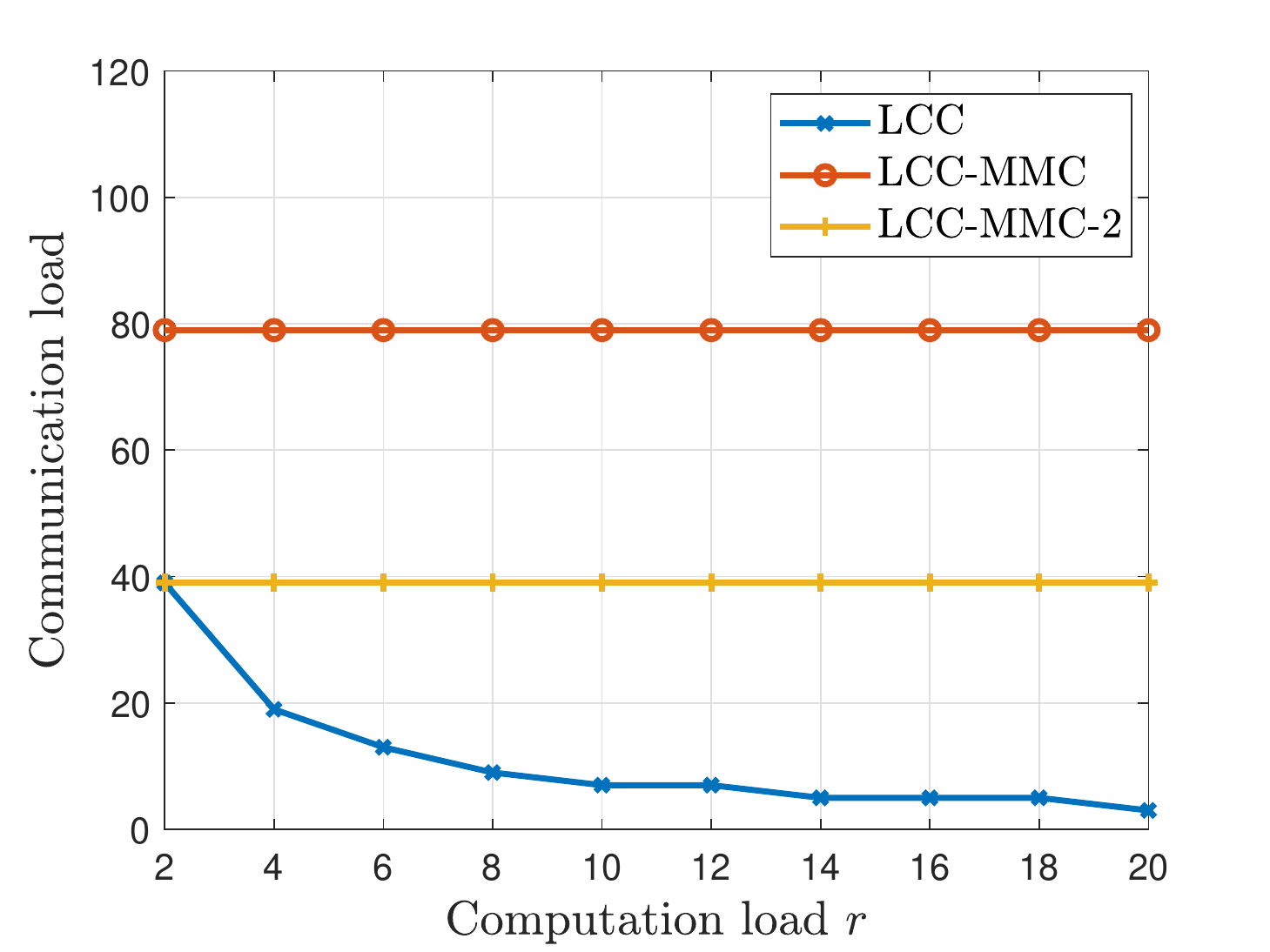}
        \caption{Communication load vs computation load }
				\label{comm2}
        \end{subfigure}
				\caption{Per iteration completion time and communication load statistics. }
		\label{avg2}
\end{figure*}

\begin{remark}
An important aspect of the average per-iteration completion time that is ignored here, and by other works in the literature, is the decoding complexity at the AS. Among these three schemes, UC-MMC has the lowest decoding complexity, while LCC-MMC has the highest. However, as discussed in Section 2, the number of transmissions as well as the decoding complexity  can be reduced via increasing the number of polynomials used in the decoding process. To illustrate this, we consider a different implementation of the LCC-MMC scheme, where two polynomials are used in the encoding part, denoted by LCC-MMC-2. In this scheme, for given $r$, the coded inputs correspond to the evaluation of two polynomials with each degree $N-1$, at $r/2$ different points. Each CS sends a partial result to AS after execution of two computations, which correspond to the evaluation of these two polynomials at the same point. Since two polynomials are used in the encoding, the number of transmissions is reduced approximately to half compared to LCC-MMC as illustrated in Fig. \ref{avg2}(b). Although a noticeable improvement is achieved in the communication load, we observe a relatively small increase in the average per iteration completion time as illustrated in Fig. \ref{avg2} (a).
\end{remark}

Overall, the optimal strategy highly depends on the network structure. When the completion time  is dominated by the CSs' computation time, the LCC-MMC becomes the best alternative. This might be the case when the workers represent GPUs or CPUs on the same machine. On the other hand, if the communication load is the bottleneck, then LCC becomes more attractive especially when the servers have enough storage capacity, i.e., large r. However, as we observe in Fig. \ref{avg2}, the communication load and the average per iteration completion time can be balanced via playing with the number of polynomials used in the encoding process; hence, the per iteration completion time can be reduced further without causing excessive increase in the communication load. We also note here that  it has been recently shown in \cite{communication_eff} that the communication load can be reduced further by doing consecutive matrix multiplications at the CSs over several iterations without communicating with AS, and  then sending higher degree coded matrix multiplication results to the AS. In the end, the AS interpolates a polynomial with a higher degree, which requires a larger non-straggling threshold compared to LCC, but with a benefit of drastically reduced communication load. However, we note that implementation of the proposed strategy is limited by the number of CSs since the non-straggling threshold can not be larger than the number CSs. 

\begin{figure*}
    \centering
         \begin{subfigure}[b]{0.47\textwidth}
        \includegraphics[scale=0.6]{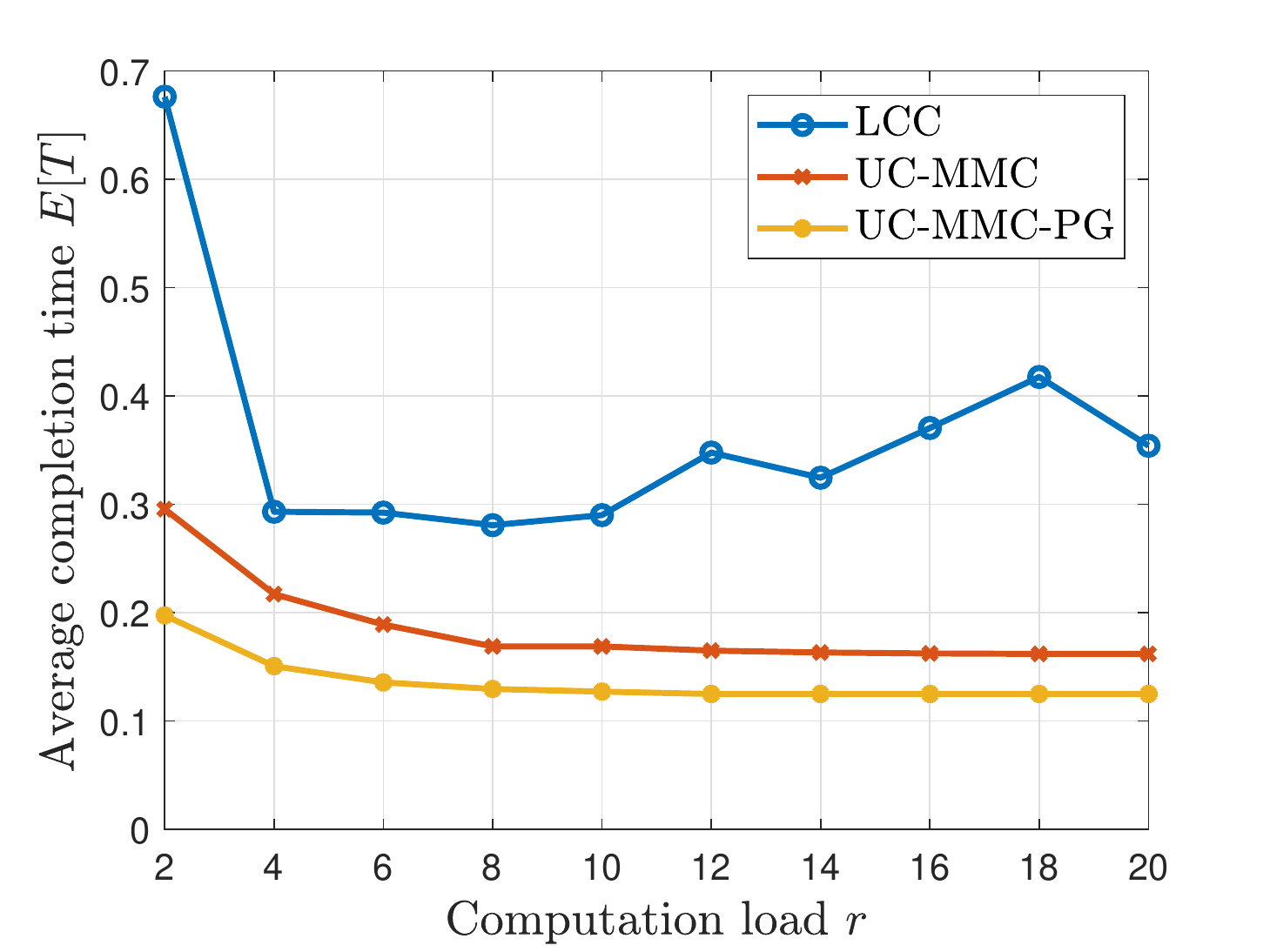}
        \caption{Average completion time vs computation load}
				\label{comp3}
    \end{subfigure}
    \begin{subfigure}[b]{0.47\textwidth}
        \includegraphics[scale=0.6]{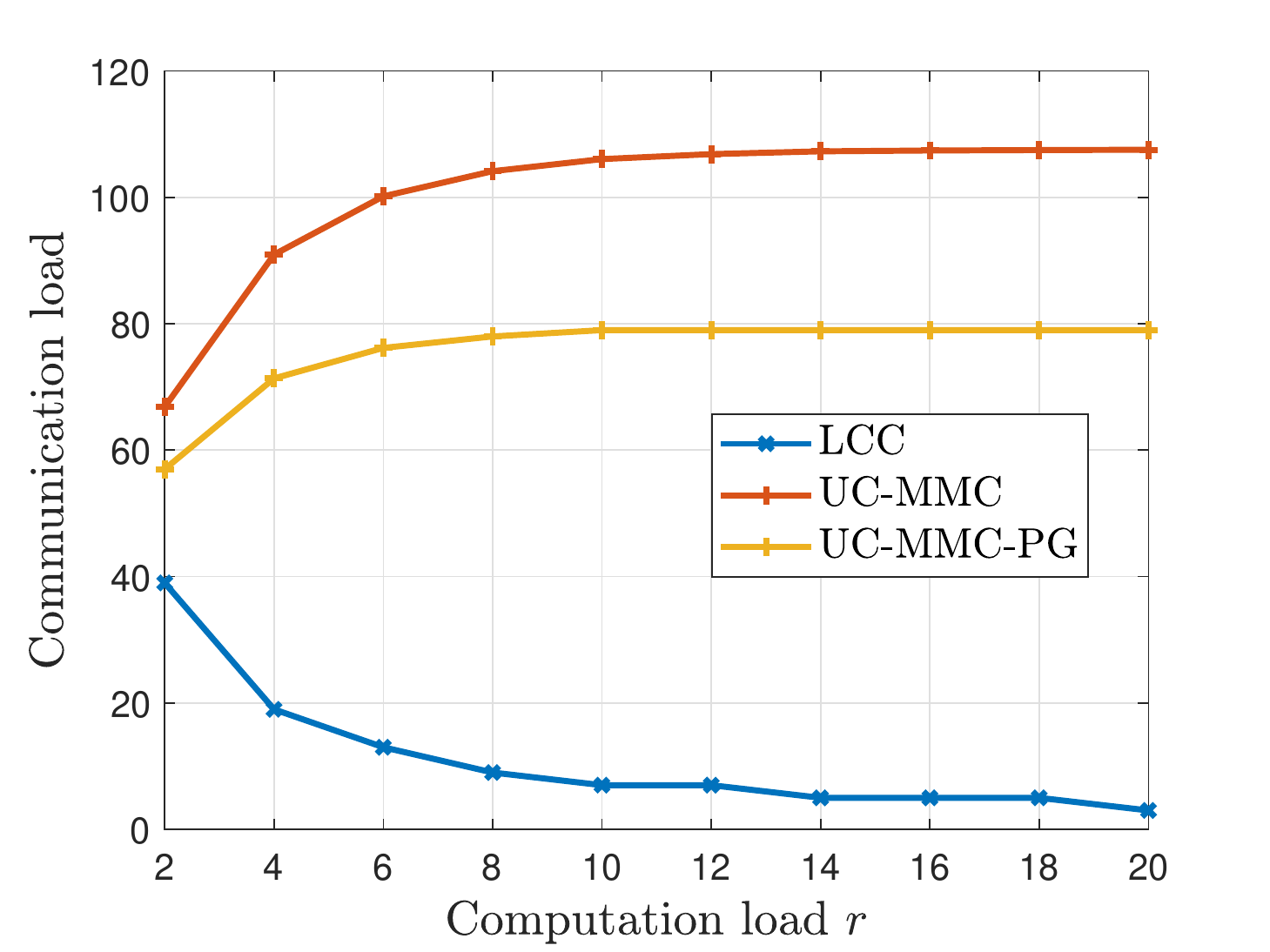}
        \caption{Communication load vs computation load }
				\label{comm3}
        \end{subfigure}
				\caption{Per iteration completion time and communication load statistics. }
		\label{avg3}
\end{figure*}

We also observe that when the 
CSs have a small storage capacity, i.e., small $r$, UC-MMC has the lowest per iteration completion time. Moreover, when the decoding complexity is taken into account, UC-MMC can be preferable to  coded computation schemes. Another advantage of the UC-MMC scheme is its applicability to K-batch SGD. The coded computation approaches are designed to obtain the full gradient; hence, at each iteration, they  wait until they can recover all the gradient values. However, in the K-batch stochastic gradient descent approach the parameter vector $\boldsymbol{\theta}_t$ is updated when any $K$ gradient values, corresponding to different batches (data points), are available at the AS. Using gradients corresponding to $K$ data points, instead of the full gradient, the per iteration completion time can be reduced. To this end, we consider a partial gradient scheme with multi-message communication, UC-MMC-PG, with $5\%$ tolerance, i.e., $K=N\times0.95$. We plot the average completion time and communication loads for different values of $r$ in Fig. \ref{avg3}. The results show that when $r$ is  small, UC-MMC-PG  can reduce the average completion time up to $70\%$ compared to LCC, and up to $33\%$ compared to UC-MMC; while only two gradient values are missing at each iteration. In addition to the improvement in average completion time, the UC-MMC-PG scheme can also reduce the communication load as shown in Fig. \ref{avg3}(b). We remark that, in the K-batch approach the gradient used for each update is less accurate compared to the full-gradient approach; however, since the parameter vector $\boldsymbol{\theta}_t$ is updated over many iterations, K-batch approach may converge to the optimal value faster than the full-gradient approach.

\section{Conclusions and Future  Directions}
We have introduced novel coded and uncoded DGD schemes when  multi-message communication is allowed from each server at each iteration of the DGD algorithm. We first provided a closed-form expression for the per iteration completion time statistics of these schemes, and verified our results with   Monte Carlo simulations. Then, we compared  these schemes with other DGD schemes in the literature in terms of the average computation and communication loads incurred. We have observed that allowing multiple messages to be conveyed from each CS at each GD iterations can reduce the average completion time significantly at the expense of an increase in the average communication load. Depending on the network structure, communication protocol employed, and computation capabilities of the CSs, we have proposed a generalized coded DGD scheme that can provide a balance between the communication load and the completion time. We also observed that UCUC with simple circular shift can be  more efficient  compared to coded computation approaches when the servers have limited storage capacity. We emphasize that, despite benefits of coded computation in reducing the computation time, their relevance in practical big data problems is questionable due to the need to jointly transform the whole data set, which may not even be possible to store in a single server. As a future extension of this work we will analyze the overall performance of these schemes in a practical setup  for a more realistic comparison. 



\ifCLASSOPTIONcaptionsoff
  \newpage
\fi



\bibliographystyle{IEEEtran}
\bibliography{IEEEabrv,SGDref}

\end{document}